# Kafka versus RabbitMQ*

PHILIPPE DOBBELAERE, Nokia Bell Labs
KYUMARS SHEYKH ESMAILI, Nokia Bell Labs

Publish/subscribe is a distributed interaction paradigm well adapted to the deployment of scalable and loosely coupled systems.

Apache Kafka and RabbitMQ are two popular open-source and commercially-supported pub/sub systems that have been around for almost a decade and have seen wide adoption. Given the popularity of these two systems and the fact that both are branded as pub/sub systems, two frequently asked questions in the relevant online forums are: how do they compare against each other and which one to use?

In this paper, we frame the arguments in a holistic approach by establishing a common comparison framework based on the core functionalities of pub/sub systems. Using this framework, we then venture into a qualitative and quantitative (i.e. empirical) comparison of the common features of the two systems. Additionally, we also highlight the distinct features that each of these systems has. After enumerating a set of use cases that are best suited for RabbitMQ or Kafka, we try to guide the reader through a determination table to choose the best architecture given his/her particular set of requirements.

## 1 INTRODUCTION

The Internet has considerably changed the scale of distributed systems. Distributed systems now involve thousands of entities fi!? potentially distributed all over the world fi!? whose location and behavior may greatly vary throughout the lifetime of the system. These constraints underline the need for more flexible communication models and systems that reflect the dynamic and decoupled nature of the applications. Individual point-to-point and synchronous communications lead to rigid and static applications, and make the development of dynamic large-scale applications cumbersome [15]. To reduce the burden of application designers, the glue between the different entities in such large-scale settings should rather be provided by a dedicated middleware infrastructure, based on an adequate communication scheme. The publish/subscribe interaction scheme provides the loosely coupled form of interaction required in such large scale settings [15].

Apache Kafka [1] and RabbitMQ [4] are two popular open-source and commercially-supported pub/sub systems (by Confluent Inc. and Pivotal) that have been around for almost a decade and have seen wide adoption in enterprise companies.

Despite commonalities, these two systems have different histories and design goals, and distinct features. For example, they follow different architectural models: In RabbitMQ, producers publish (batches of) message(s) with a routing key to a network of exchanges where routing decisions happen, ending up in a queue where consumers can get at messages through a push (preferred) or pull mechanism. In Kafka producers publish (batches of) message(s) to a disk based append log that is topic specific. Any number of consumers can pull stored messages through an index mechanism.

Given the popularity of these two systems and the fact that both are branded as pub/sub systems, two frequently asked questions in the relevant online forums are: how do they compare against each other and which one to use?

---

While one can find several ad-hoc recommendations (some based on pointed benchmarks) on the web, we found these hard to generalize to other applications and believe they do not do justice to the systems under discussion. More specifically, (i) the bigger context of such analysis, e.g., qualitative comparison or the distinct features of each system is often overlooked, (ii) due to the fast pace of developments, in particular in the case of Kafka, some of the reported results are stale and no longer valid, and (iii) it is difficult to compare results across different experiments.

In this paper, we frame the arguments in a more holistic approach. More concretely, we start (in Section 2) with a brief description of the core functionalities of publish/subscribe systems as well as the common quality-of-service guarantees they provide. We then (in Section 3) give a high-level description of both Apache Kafka and RabbitMQ. Based on the framework established in Section 2, we venture into a qualitative (in Section 4) and quantitative (in Section 5) comparison of the common features of the two systems. In addition to the common features, we list the important features that are unique to either of the two in Section 6. We next enumerate a number of use case classes that are best-suited for Kafka or RabbitMQ, as well as propose options for a combined use of the two systems in Section 7. There, we also propose a determination table to help choose the best architecture when given a particular set of requirements. Finally, we conclude the paper in Section 8.

## 2 BACKGROUND: PUB/SUB SYSTEMS

In this section we highlight the main concepts of the publish/subscribe paradigm, its required and desired guarantees as well as some of its realizations out there.

The primary purpose of this section is to establish a common framework/language that will be used in the rest of the paper. Knowledgeable readers may skip it.

### 2.1 Core Functionalities

Publish/subscribe is a distributed interaction paradigm well adapted to the deployment of scalable and loosely coupled systems.

Decoupling the publishers and subscribers is arguably the most fundamental functionality of a pub/sub system. Eugster et al. [15] have decomposed the decoupling that the pub/sub coordination scheme provides along the following three dimensions:

(1) **Entity decoupling**: publishers and consumers do not need to be aware of each other. The pub/sub infrastructure terminates the interaction in the middle.
(2) **Time decoupling**: The interacting parties do not need to be actively participating in the interaction, or even stronger, switched on, at the same time.
(3) **Synchronization decoupling**: the interaction between either producer or consumer and the pub/sub infrastructure does not synchronously need to block the producer or consumer execution threads, allowing maximum usage of processor resources at producers and consumers alike.

Another core functionality of pub/sub systems is **routing logic** (also known as subscription model) which decides if and where a packet that is coming from a producer will end up at a consumer. The different ways of specifying the events of interest have led to several subscription schemes, that balance flexibility against performance. The two main types of routing logic are the following:

- A **topic-based** subscription is characterized by the publisher statically tagging the message with a set of topics, that can then be used very efficiently in the filtering operation that decides which message goes to which consumer. Most systems allow topic names to contain wildcards, and topic names can have hierarchy to enhance the filtering capabilities, at the expense of higher processor load.
- A **content-based** subscription does not need the producer to explicitly tag the message with routing context. All data and metadata fields of the message can be used in the filtering condition. Consumers subscribe to selective events by specifying filters using a subscription language. The filters define constraints, usually in the form of name-value pairs of properties and basic comparison operators, which identify valid events. Constraints can be logically combined (and, or, etc.) to form complex subscription patterns. Evaluating these complex filters comes at a high processing cost.

## 2.2 Quality-of-Service Guarantees

In addition to the aforementioned core functionalities of pub/sub systems, they are also defined by a relatively large set of required and desired guarantees that are generally referred to as Quality-of-Service (QoS) guarantees [9, 11, 15].

For sake of simplicity, we have grouped the most important pub/sub QoS guarantees into five separate categories and will explain them in the following sections.

It should be noted that an important assumption in this section is the distributed nature of modern pub/sub systems. Distribution is necessary (but not sufficient) to bring scalability. However, it brings a number of technical problems that make the design and implementation of distributed storage, indexing and computing a delicate issue [7].

*2.2.1 Correctness.* As proposed in [29], correctness behavior can be defined using three primitives: no-loss, no-duplication, no-disorder. Building upon these primitives, the following two criteria are relevant in pub/sub systems:

- *Delivery Guarantees*, the three common variants are:
  - **at most once** (aka "best effort"; guarantees no-duplicates): in this mode, under normal operating conditions, packets will be delivered, but during failure packet loss might occur. Trying to do better than this will always cost system resources, so this mode has the best throughput.
  - **at least once** (guarantees no-loss): in this mode, the system will make sure that no packets get lost. Recovery from failures might cause duplicate packets to be sent, possibly out-of-order.
  - **exactly once** (guarantees no-loss and no-duplicates): this requires an expensive end-to-end two phase commit.
- *Ordering Guarantees*, the three common variants here are:
  - **no ordering**: absence of ordering guarantees is an ideal case for performance.
  - **partitioned ordering**: in this mode, a single partition can be defined for each message flow that needs to be consumed in-order. While more expensive than the previous group, it can possibly have high performance implementations.
  - **global order**: due to the synchronization overhead, imposing a global ordering guarantee across different channels requires significant additional resources and can severely penalize performance.

*2.2.2 Reliability.* denotes the ability of a distributed system to deliver it's services even when one or several of it's software of hardware components fail.

It definitely constitutes one of the main expected advantages of a distributed solution, based on the assumption that a participating machine affected by a failure can always be replaced by another one, and not prevent the completion of a requested task.

An immediate and obvious consequence is that reliability relies on redundancy of software components, network connections and data. Clearly, this has a cost.

*2.2.3 Availability.* Availability is the capacity of a system to maximize its uptime. Note that this implicitly assumes that the system is already reliable: failures can be detected and repair actions initiated.

*2.2.4 Transactions.* In messaging systems, transactions are used to group messages into atomic units: either a complete sequence of messages is sent (received), or none of them is. For instance, a producer that publishes several semantically related messages may not want consumers to see a partial (inconsistent) sequence of messages if it fails during emission.

Similarly, a mission-critical application may want to consume one or several messages, process them, and only then commit the transaction. If the consumer fails before committing, all messages are still available for reprocessing after recovery.

*2.2.5 Scalability.* The concept of scalability refers to the ability of a system to continuously evolve in order to support a growing amount of tasks. In the case of pub/sub systems, scalability can have various dimensions e.g., consumers/producers, topics and messages.

*2.2.6 Efficiency.* Two common measures of efficiency are the *latency* (or response time), and the *throughput* (or bandwidth).

**Latency.** In any transport architecture, latency of a packet/message is determined by the serial pipeline (i.e., sequence of processing steps) that it passes through.

Latency for any transport architecture can be defined as the time delay incurred by a packet from the moment it enters to the moment it exits part of an architecture. In this paper, we will primarily focus on latency inside a network node. When the transport architecture is distributed over multiple nodes that are not collocated additional network latencies will need to be added. Examples of the latter case are applications that span end-to-end network topologies all the way from edge to core networks, or involve geographical aggregation.

In pub/sub system, typically the main latency contributors are as follows:

- compute cycles needed for packet metadata handling (validating, routing, ...), typically not packet size dependent
- compute cycles needed for packet copy, typically packet size dependent
- storage access latency (write versus read, DRAM versus disk, sequential disk access versus random disk access)
- persistence and ordering overhead in case of "at least once" and/or ordered delivery needs to be guaranteed
- dequeueing latency (at consumer speed) due to the FIFO behavior of a queue which is not empty

Latency can only be reduced by pipelining the packet transport over resources that can work concurrently on the same packet in a series architecture (multiple processing cores, master DMA engines in case of disk or network access,...) . It is not influenced by scaling out resources in parallel.

**Throughput.** Throughput of a transport architecture is the number of packets (or alternatively, bytes) per time unit that can be transported between producers and consumers. Contrary to latency, throughput can easily be enhanced by adding additional resources in parallel.

For a simple pipeline throughput and latency are inversely proportional.

It is important to point out that both efficiency and scalability often conflict with other desirable guarantees [15]. For instance, highly expressive and selective subscriptions require complex and expensive filtering and routing algorithms, and thus limit scalability. Similarly, strong availability and delivery guarantees entail considerable overheads, due to cost attached to persistence and replication and the fact that missed events must be detected and retransmitted.

## 2.3 Realizations

A large number of frameworks and libraries can be categorized as having pub/sub messaging functionality. One approach to categorize them is to locate them on a complexity spectrum that starts with lightweight systems with fewer features and ends with complex systems that offer a rich set of functionalities.

At the lightweight side of the spectrum, we find ZeroMQ, Finagle, Apache Kafka, etc. Heavier examples include the Java Message Service (JMS) implementations such as ActiveMQ, JBOSS Messaging, Glassfish, etc. AMQP 0.9, the popular and standardized pub/sub protocol has several implementations such as RabbitMQ, Qpid, HornetQ, etc. Even more complex and feature-rich are distributed RPC frameworks that include pub/sub, e.g., MuleESB, Apache ServiceMix, JBossESB, etc.

## 3 HIGH-LEVEL DESCRIPTION

In this section we give a brief description of the Apache Kafka and RabbitMQ systems. In particular, we look at the history/context of their creation, their main design goals, as well as some notable implementation and optimization details about them. Each of these aspects can help us gain further insights about these systems and hence better explain their differences.

### 3.1 Apache Kafka

Kafka was originally built at LinkedIn as its centralized event pipelining platform, replacing a disparate set of point-to-point integration systems [18].

The Kafka team had initially explored a number of alternatives, most notably ActiveMQ, a popular messaging system based on JMS. However, in production tests it ran into two significant problems: (i) if the queue backed up beyond what could be kept in memory, performance would severely degrade due to heavy amounts of random I/O, (ii) having multiple consumers required duplicating the data for each consumer in a separate queue.

The conclusion was that messaging systems target low-latency settings rather than the high-volume scale-out deployment that was required at LinkedIn. Consequently, they decide to build a piece of custom infrastructure meant to provide efficient persistence, handle long consumer backlogs and batch consumers, support multiple consumers with low overhead, and explicitly support distributed consumption while retaining the clean real-time messaging abstraction of messaging systems.

The resulting system is a scalable publish-subscribe messaging system designed around a distributed commit log [33]. High-throughput is one advantage of the design of log aggregation systems over most messaging systems [18]. Data is written to a set of log files with no immediate flush to disk, allowing very efficient I/O patterns.

Figure 1 shows the high-level architecture of Kafka. Producers send messages to a Kafka topic that holds a feed of all messages of that topic. Each topic is spread over a cluster of Kafka brokers, with each broker hosting zero or more partitions of each topic. Each partition is an ordered write-ahead log of messages that are persisted to disk. All topics are available for reading by any number of consumers, and additional consumers have very low overhead.

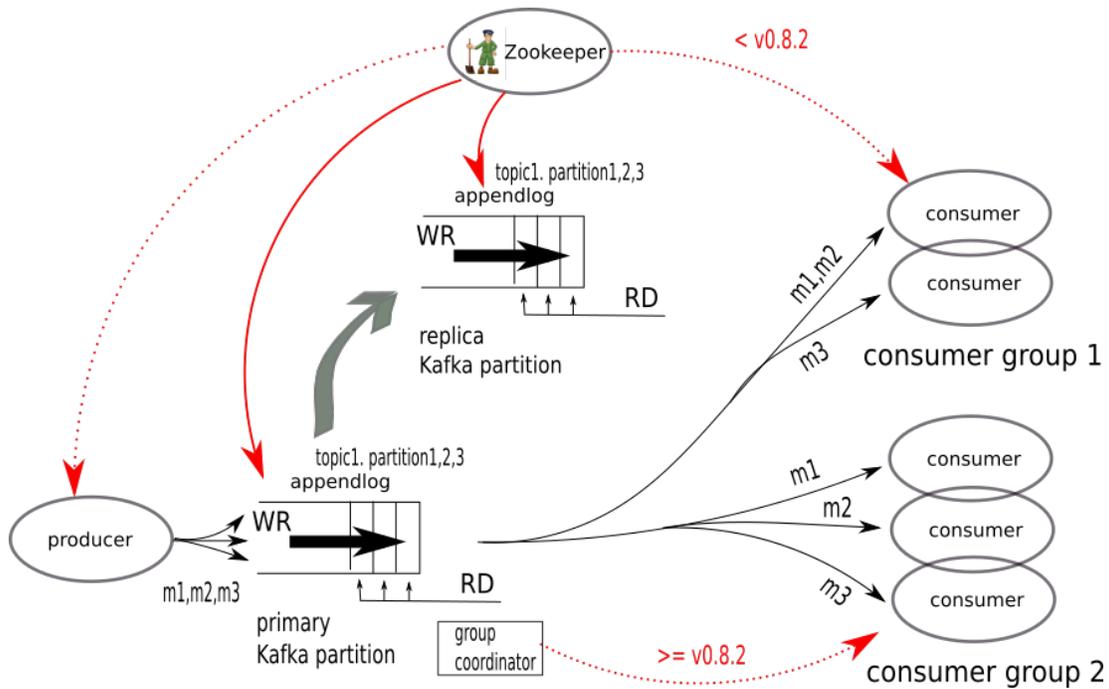

Fig. 1. Kafka Architecture

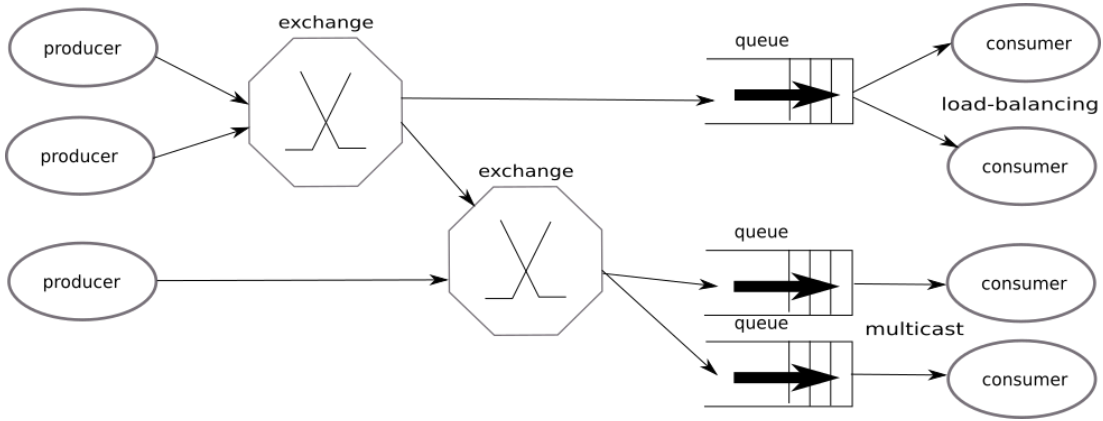

Fig. 2. RabbitMQ (AMQP) Architecture

Kafka has a very simple storage layout. Each partition of a topic corresponds to a logical log. Physically, a log is implemented as a set of segment files of approximately the same size (e.g., 1GB). Every time a producer publishes a message to a partition, the broker simply appends the message to the last segment file.

Compared to the traditional pub/sub systems, the notion of a consumer in Kafka is generalized to be a group of co-operating processes running as a cluster. Each message in the topic is delivered to one consumer in each of these consumer groups. As a result the partition is the unit of parallelism of the topic and controls the maximum parallelism of the consumers. Furthermore, because each partition has a sole consumer within its group, the consuming process can lazily record its own position, rather than marking each message immediately, which is essential for performance. If the process crashes before the position is recorded it will just reprocess a small number of messages, giving at-least-once delivery semantics.

| Batching Type | Improvement | Default Threshold |
|---|---:|---:|
| Producer | 3.2x | 200 messages or 30 seconds |
| Broker | 51.7x | 50000 messages or 30 seconds |
| Consumer | 20.1x | 1 megabyte |

Table 1. Improvements due to batching in Kafka [18].

Message producers balance load over brokers and sub-partitions either at random or using some application-supplied key to hash messages over broker partitions. This key-based partitioning has two uses. First the delivery of data within a Kafka partition is ordered but no guarantee of order is given between partitions. Consequently, to avoid requiring a global order over messages, feeds that have a requirement for ordering need to be partitioned by some key within which the ordering will be maintained. The second use is in support of a routing/filtering feature: consumers can partition a data stream by a key such as the user id, and perform simple in-memory session analysis distributed across multiple processes relying on the assumption that all activity for a particular user will be sticky to one of those consumer processes. Without this guarantee distributed message processors would be forced to materialize all aggregate state into a shared storage system, likely incurring an expensive look-up round-trip per message.

Finally, it is worth noting that originally Kafka relied heavily on Apache Zookeeper [19] for the implementation of its control plane logic, but the Zookeeper reliance is trimmed down with every release. Around v0.8.2, consumer management was transferred from Zookeeper to a coordinator inside the broker. Still managed by Zookeeper are controller and cluster management, topic and partition management, in-sync data replication and static configurations like quotas and ACLs.

In summary, to meet the high-throughput requirements, Kafka has departed from the classic principles of messaging systems in a few ways:

- It partitions up data so that production, brokering, and consumption of data are all handled by clusters of machines that can be scaled incrementally as load increases. Kafka guarantees that messages from a single partition are delivered to a consumer in order. However, there is no guarantee on the ordering of messages coming from different partitions.
- Messages are not "popped" from the log, but can be replayed by the consumers (e.g. when handling consumer application errors)
- Additionally, reader state is kept only by the consumers, implying that message deletion can only be based on a manually-tuned retention policy, expressed either in message count or message age.

Furthermore, it also applies a number of very effective optimization techniques, most notably:

- As shown in Table 1, it uses batching at all stages of the pipeline (production, brokering, and consumption) with significant throughput improvements.
- It relies on persistent data structures and OS page cache. The operating systemfis read-ahead strategy is very effective for optimizing the linear read pattern of consumers which sequentially consume chunks of log files. The buffering of writes naturally populates this cache when a message is added to the log, and this in combination with the fact that most consumers are not far behind, means a very high cache hit ratio making reads nearly free in terms of disk I/O.

## 3.2 RabbitMQ

RabbitMQ is primarily known and used as an efficient and scalable implementation of the Advanced Message Queuing Protocol (**AMQP**). Hence, below we first give a short introduction of AMQP, and then briefly explain the RabbitMQ implementation (and extensions) of it.

*3.2.1 AMQP.* AMQP was born out the need for interoperability of different *asynchronous* messaging middlewares. More concretely, while various middleware standards existed for synchronous messaging (e.g., IIOP, RMI, SOAP, etc), the same did not hold true in the world of asynchronous messaging, however, in which several proprietary products exist and use their own closed protocols (e.g. IBM Websphere MQ and Microsoft Message Queuing) [32]. Java Message Service (JMS) specification was arguably the best-known standard in the asynchronous messaging world. However, it is merely an interface standard and does not specify a standard protocol. Furthermore, JMS was limited to Java, which is only one viable implementation technology within the messaging middleware domain.

What is now known as AMQP originated in 2003 at JPMorgan Chase. From the beginning AMQP was conceived as a co-operative open effort. JPMorgan Chase partnered with Red Hat to create Apache Qpid. Independently, RabbitMQ was developed in Erlang by Rabbit Technologies.

Around 2011, the AMQP standard bifurcated away from the widely-adopted v0.9.1 (a slight variation of version 0.9 [31]) functionality with the creation of AMQP 1.0. Compared to Java Message Service (JMS), which just defines an API, AMQP defines a binary protocol implementation that guarantees interoperability between different parties implementing the protocol independently.

The design of AMQP has been driven by stringent performance, scalability and reliability requirements from the finance community. However, its use goes far beyond the the financial services industry and has general applicability to a broad range of middleware problems.

As shown in Figure 2, AMQP takes a modular approach, dividing the message brokering task between **exchanges** and message **queues** [32]:

- An exchange is essentially a router that accepts incoming messages from applications and, based on a set of rules or criteria, decides which queues to route the messages to.
- A message queue stores messages and sends them to message consumers. The storage mediumfis durability is entirely up to the message queue implementation –message queues typically store messages on disk until they can be delivered– but queues that store messages purely in memory are also possible.

Joining together exchanges and message queues are **bindings**, which specify the rules and criteria by which exchanges route messages. Specifically, applications create bindings and associate them with message queues, thereby determining the messages that exchanges deliver to each queue.

AMQP assumes a stream-based transport (normally TCP) underneath it. It transmits sequential frames over channels, such that multiple channels can share a single TCP connection. Each individual frame contains its channel number, and frames are preceded by their sizes to allow the receiver to efficiently delineate them. AMQP is a binary protocol.

In a multi-threaded environment, individual threads are typically assigned their own channel.

*3.2.2 RabbitMQ Implementation and Extensions of AMQP.* RabbitMQ, by default, supports AMQP 0.9.1 and can support AMQP 1.0 through a plugin.

RabbitMQ goes beyond the AMQP guarantees in a number of aspects: it has more efficient acknowledgment mechanism for the publishers, has better-defined transactional behavior, has better support for asynchronous batch transfer, supports a degree of coupling between producers and consumers (i.e the flow control). For a detailed list of extensions, see [2].

RabbitMQ is implemented in Erlang, which implies it uses the Actor Model as communication primitive between lightweight Erlang processes. It therefore profits from the Erlang Open Telecom Platform (OTP) infrastructure which greatly facilitates the creation and management of high-availability architectures. Erlang and the actor model are the prime reasons for the scalability capabilities of RabbitMQ in terms of number of topics and queues, and bring clustering capabilities at a very low design overhead.

Compared to Kafka, RabbitMQ is much closer to the classic messaging systems. More specifically, RabbitMQ: (i) takes care of most of the consumption bookkeeping, (ii) its main design goal is to handle messages in DRAM memory, (iii) the queue logic is optimized for empty-or-nearly-empty queues and the performance degrades significantly if messages are allowed to accumulate [1]

[5].

## 4 COMMON FEATURES: QUALITATIVE COMPARISON

In this section we give a qualitative comparison of Kafka and RabbitMQ across a number of common pub/sub features.

It should be noted that for the sake of simplicity, we only consider recent stable releases of the two systems (i.e. **Kafka 0.10** and **RabbitMQ 3.5**).

### 4.1 Time Decoupling

Both systems can be used to buffer a large batch of messages that needs to be consumed at a later time or at a much lower rate than it is produced.

To this end, RabbitMQ will store the messages in DRAM as long as possible, but once the available DRAM is completely consumed, RabbitMQ will start storing messages on disk without having a copy available in DRAM, which will severely impact performance.

Kafka, on the other hand, was specifically designed with the various consumption rates requirement in mind and hence is much better positioned to handle a wider scale of time decoupling.

### 4.2 Routing Logic

RabbitMQ inherits the routing logic of AMQP and hence can be very sophisticated. Stock RabbitMQ already provides for a number of different exchange types, most notably:

- a very flexible topic-based exchange (of type *topic*) that supports multipart "a.b.c" topic-based routing with wildcard support ("*" for one part and "#" for an arbitrary number of parts),
- a content-based exchange (of type *header*).

Since RabbitMQ provides an API to create additional exchanges, routing logic can be anything you need. For example, the RabbitMQ community has provided additional exchange definitions, most importantly support for load balancing [3, 28].

Another relevant and useful feature in RabbitMQ is *Alternate Exchange* which allows clients to handle messages that an exchange was unable to route (i.e. either because there were no bound queues our no matching bindings).

With Kafka, the choice is more limited: it supports a basic form of topic-based routing. More specifically, the producer controls which partition it publishes messages to. This can be done at random (i.e. load balancing) or by some partitioning function by allowing the user to specify a partition-by key and using this to hash to a partition. The partition function can be overridden by the user.

---

[1] Probably due to cross-fertilization from Kafka, RabbitMQ introduced the concept of Lazy Queues in v3.6. Lazy Queues store messages immediately to disk, and only read them back in memory when consumers start reading from the queue.

### 4.3 Delivery Guarantees

RabbitMQ and Kafka differ in their notion of at least once semantics. Since individual packets from a batch can fail, recovery from failures can have impact on the order of packets. Depending on the application, order might be important, so it makes sense to split this up in

(1) at least once **without** order conservation: Kafka cannot preserve order when sending to multiple partitions.
(2) at least once **with** order conservation: RabbitMQ sorts messages when writing them to queue structures, meaning that lost messages can be correctly delivered in order without the need to resend the full batch that lost 1 or more messages. Kafka will preserve order under conditions specified in Section 4.4.

It should be noted that using standard AMQP 0.9.1, the only way to guarantee that a message is not lost is by using transactions which are unnecessarily heavyweight and drastically decrease throughput. To remedy this, in RabbitMQ a confirmation mechanism was introduced which mimics the consumer acknowledgments mechanism already present in the protocol.

Guaranteeing that a packet gets delivered involves the concept of "ownership transfer" between the different components of the architecture. A guarantee is not absolute: we introduce the notion of failure probability over time and the failure rate $\lambda$ of individual components and of the complete packet transfer chain. Failure probability and rate can be reduced by providing replication.

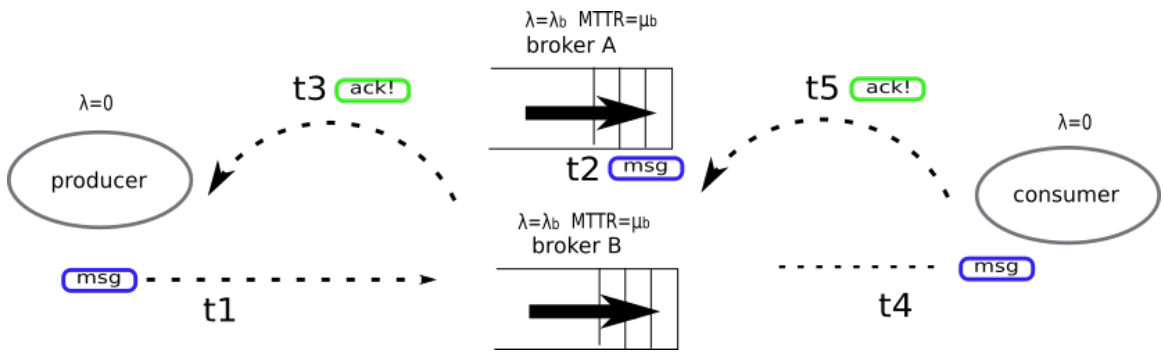

Fig. 3. reliable transfer

In the following, producer and consumer failures are out of scope ( we assume $\lambda = 0$ ).

The scenarios for RabbitMQ and Kafka mainly digress in the generation of publisher confirms, the consumer interaction and message deletion aspects.

- t1, the producer owns a message to be forwarded and delivers it to RabbitMQ/Kafka.
- t2, RabbitMQ "handles" the message; the actual logic of this handling is case-specific:
  (1) for unroutable messages, the broker will issue a confirm once the exchange verifies a message would not route to any queue,
  (2) for routable messages, the confirmation is issued once the message has been accepted by all the queues,
  (3) for persistent messages routed to durable queues, this means persisting to disk, and
  (4) for mirrored queues, this means that all mirrors have accepted the message
  Kafka appends the message to the relevant partition of the append log on the master broker node A and potentially on a redundant broker node B

- t3, a coordinated ACK from node A (and if applicable, B) is sent to the producer - ownership now moved to RabbitMQ/Kafka and the producer can delete the message
- t4, the consumer gets the message from RabbitMQ/Kafka
- t5 [RabbitMQ specific] the consumer sends an ACK to node A (and if applicable, B) - ownership now moved to the consumer and the broker can delete the message. Note that typically every consumer will read from a dedicated queue, so the broker will keep ownership of messages that need to go to multiple consumers if all ACKS are not yet received.
- t5 [Kafka specific] Kafka is not keeping state, so has no way of understanding ownership moved to the consumer. It will keep hold of the message until a configured timeout expires (typically several days).

RabbitMQ improves on AMQP and offers the possibility to publish batches of messages with individual ACK/NACK replies indicating that the message safely made it to disk (i.e. fsynced[2] to the storage medium).

The acknowledgment behavior of Kafka (request.required.acks) can be chosen as 0 for best effort, 1 to signal the producer when the leader has received the packet but did not commit it to disk (meaningful while running under replication since otherwise packet could get lost), or −1 to signal the producer when a quorum has received the packet but did not commit it to disk (should not be a problem unless all replicas run in the same environment, which implies they could all go down at once caused by e.g. a power failure).

While running without replication, Kafka in its default configuration does not wait with sending ACKs until an fsync has occurred and therefore messages might be lost in the event of failure. This can be changed by configuration, at the expense of a reduction in throughput.

### 4.4 Ordering Guarantees

RabbitMQ will conserve order for flows[3] using a single AMQP channel. It also reorders retransmitted packets inside its queue logic so that a consumer does not need to resequence buffers. This implies that if a load-balancer would be used in front of RabbitMQ (e.g. to reach the scalability of what can be accomplished inside Kafka with partitions), packets that leave the load-balancer on different channels will have no ordering relation anymore.

Kafka will conserve order only inside a partition. Furthermore, within a partition, Kafka guarantees that a batch of messages either all pass or all fail together. However, to conserve inter-batch order, the producer needs to guarantee that at most 1 produce request is outstanding, which will impact maximum performance.

### 4.5 Availability

Both RabbitMQ and Kafka provide availability via replication.

RabbitMQ Clusters can be configured to replicate all the exchange and binding information. However, it will not automatically create mirrored queues (RabbitMQ's terminology for replicated queues) and will require explicit setting during queue creation.

For Kafka, availability requires running the system with a suitably high replication factor.

As stated by the CAP theorem [17], in any architecture based on replication, split-brain problems can arise due to fault induced network partitions. For an in-depth description of the availability

---

[2] the OS call fsync() transfers all modified data in DRAM to the disk device so that all changed information can be retrieved even after the system crashed or was restarted

[3] a sequence of messages that is to be processed without insertion, deletion or reordering

models (as well as CAP theorem analysis) of Kafka and RabbitMQ see the corresponding episodes in the Jepsen series [20, 21].

### 4.6 Transactions

AMQP transactions only apply to publishes and acks. RabbitMQ has additionally made rejection transactional. On the consuming side, the acknowledgments are transactional, not the consuming of the messages themselves. AMQP guarantees atomicity only when transactions involve a single queue. RabbitMQ provides no atomicity guarantees even in case of transactions involving just a single queue, e.g. a fault during *commit* can result in a sub-set of the transaction's publishes appearing in the queue after a broker restart. Note that these are not transactions in the strict ACID sense, since some interaction with the publisher or consumer is required. Take e.g. a producer publishing a batch. If any of the messages fails, the producer gets the chance to republish these messages, and RabbitMQ will insert them in the queue in order. After which the publisher is notified that the failing messages did make it and can consider the transaction complete.

Kafka currently does not support transactions. However, a proposal to extend it with this feature in the future releases has recently been adopted.

### 4.7 Multicast

Applications often need to send the same information to multiple destinations.

RabbitMQ supports multicast by providing a dedicated queue per individual consumer. As a result, the only impact on the system is that there is an increased number of bindings to support these individual queues. RabbitMQ has a view of which consumers have already taken ownership of each message, so can easily decide when it can flush the message from its system. In fan-out cases, RabbitMQ keeps per-queue indexes and metadata but only one copy of the message bodies for all queues.

In Kafka, only one copy of messages within a topic is maintained in the brokers (in non-replicated settings); however, the multicast logic is handled completely at the consumer side. Each consumer can fetch messages out of Kafka based on the message index. Kafka does not know when all consumers have taken ownership of the message, so it simply keeps the message for a configurable amount of time or size.

### 4.8 Dynamic Scaling

For RabbitMQ, adding additional nodes to running clusters or removing a node from a cluster is well supported. These additional nodes will be able to become master for newly created queues, and will accept channels allowing to publish to any exchange or consume from any queue, but cannot be used to re-distribute master queue assignments of existing queues without manual intervention. Adding nodes in a RabbitMQ cluster is transparent for consumers - these still preferably consume from the master queue, although consuming from any other cluster node works, at the expense of additional internal networking load since the packets reside on the master queue.

In Kafka, upon adding new nodes to the cluster, the user can decide to move existing partitions to the new node. In that case, a new replica is created on the new node and once it has caught up, the old replica of the partition can be deleted. This can be done online while the consumers are consuming. Adding nodes to a Kafka cluster is not transparent for consumers, since there needs to be a mapping from partitions to consumers in a consumer group. Removing a node can be done by first redistributing the partitions on that node to the remaining nodes.

## 5 COMMON FEATURES: QUANTITATIVE COMPARISON

In this section we use empirical methods to quantitatively compare the efficiency/performance of RabbitMQ and Kafka. Throughout this section, we base our arguments predominantly on our own experimental results. However, in a few cases where the required infrastructure/scenario is not easily replicable, we refer to existing results reported by others.

As explained earlier in Section 2.2.6, efficiency is primarily measured in terms of latency and throughput. Hence, we organize the content of this section accordingly: latency results are discussed in Section 5.1 and throughput results in Section 5.2.

In addition to the system and efficiency measures aspects, we include two other important dimensions in our experiments: (i) delivery guarantees, i.e. at most once vs at least once, (ii) availability, i.e., replicated queues vs non-replicated queues. As discussed in Section 2, these have important implications for efficiency.

Finally, it should be noted that while we have designed and conducted a wide set of experiments, a thorough empirical study of these two systems in beyond the scope of this paper and requires considerations of how these systems are used in the larger application architecture.

**Experimental setup.** Our experiments where conducted on a Linux server with 24 cores (Intel Xeon X5660 @ 2.80GHz) and 12GB of DRAM running a 3.11 kernel. The hard disk used was a WD1003FBYX-01Y7B0 running at 7200 rpm. Note that while use of multiple machines can make it easier to increase bandwidth to disk, but it introduces network layer effects that make it harder to define the system under test.

Both for Kafka and RabbitMQ, we used the test tools provided by the respective distributions [4]. The versions of RabbitMQ and Kafka used in the experiments were **3.5.3** and **0.10.0.1**, respectively. It should be noted that all our Kafka experiments have been carried out using the default configuration. As nicely laid out in this recent white paper [10], these default settings favor latency over throughput (most notably in the configuration parameter *linger.ms* which is by default set to 0, meaning the producer will send as soon as it has data to send).

All tests ran for 60 seconds after test setup, with results collection starting after 30 seconds. All packet generators were configured to produce maximal load. The source code of the test tools provides succinct embedded documentation of all the tunable parameters.

Whenever multiple instances of a broker were required, these were started on the same machine, effectively eliminating the majority of network latency effects.

In addition to the latency and throughput results reported below, we also monitored both the core utilization (never fully utilized, explicitly reported in some figures) and the memory consumption (never exceeded 13.4 % for RabbitMQ or 29.5% for Kafka).

Finally, since the median error across different runs were overall low (less than 10%), they are not depicted in the graphs.

### 5.1 Latency Results

We first start with the at most once mode, as it always delivers the best latency results. We then show the at least once mode results, demonstrating the cost of providing a stronger delivery guarantee.

*5.1.1 Latency in At Most Once Mode.* For RabbitMQ, the serial pipeline handling a packet mainly consists of an Erlang process terminating the protocol between the producer and the broker, a second Erlang process keeping the queue state and a third Erlang process transferring the packet

---
[4] For RabbitMQ, it's part of the Java client, under com.rabbitmq.examples.PerfTest and for Kafka, it's part of the Kafka tools under kafka.tools.

|                            | mean   | max     |
|----------------------------|--------|---------|
| with and without replication | 1–4 ms | 2–17 ms |

(a) RabbitMQ

|                     | 50 percentile | 99.9 percentile |
|---------------------|---------------|-----------------|
| without replication | 1 ms          | 15 ms           |
| with replication    | 1 ms          | 30 ms           |

(b) Kafka

Table 2. Latency Results when reading from DRAM

from the queue to the consumer. RabbitMQ latency results are optimal if the broker is allowed to have a window of outstanding unconfirmed publishes[5] (we measured with a window of 10).

For Kafka, the serial pipeline handling a packet is dominated by the storage access latency. As described in Section 3.1, Kafka directly uses the OS cache associated to the disk and, ideally, when reads occur instantly after the write, chances are high that the packet will still be available in that cache.

Our measurements of RabbitMQ and Kafka latency measurements are summarized in Table 2. Since the test tools of Kafka and RabbitMQ report different statistical summaries, in this table we have selected a subset of those that are important and semantically comparable. Our results are largely consistent with those of a similar set of experiments reported in [30].

Here are two important conclusions that can be drawn from these results: (i) Both of these systems can deliver millisecond-level low-latency guarantees. The results for Kafka seem a little better, however, as we discuss below, Kafka is operating in an ideal setting (zero cache miss) and in a more realistic setting RabbitMQ outperforms it. (ii) Replication does not drastically hamper the results. More specifically, in case of RabbitMQ the results are almost identical. For Kafka, it only appears after the median value, with a 100% increase in the 99.9 percentile.

The results reported in Table 2 are for normal (in case of Kafka, ideal) operating conditions. Below, we discuss the implications of operating beyond the normal/ideal conditions.

When RabbitMQ is running close to maximum load (an exceptional setting), the broker will start to write packets to disk to free up memory it needs for computation, effectively meaning the latency figures will rapidly deteriorate.

In case of Kafka, when consumers are slower then producers (which can be a common case), packets will have to be transferred from disk to cache before a read completes. Even with an architecture that profits from sequential disk access, the latency values will rapidly increase, not only for the slow consumer where the effect is not important but also for fast consumers that will see their cache trashed. This is demonstrated in in Figure 5 (from an experiment reported in [8]) where it shows the effect of cache miss reads when approximately 30% of the packets have to be fetched from disk, resulting in a latency increase of more than an order of magnitude.

Another factor that can severely impact Kafka latencies is the fact that Kafka runs on the JVM and large messages can cause longer garbage collection pauses as Kafka allocates large chunks. This will show up as outliers in the latency distribution. This can also negatively affect the control plane, up to the point where longer timeout values for Zookeeper (zookeeper.session.timeout.ms) need to be configured so that Kafka does not abandon the ZooKeeper session.

---

[5] https://www.rabbitmq.com/confirms.html

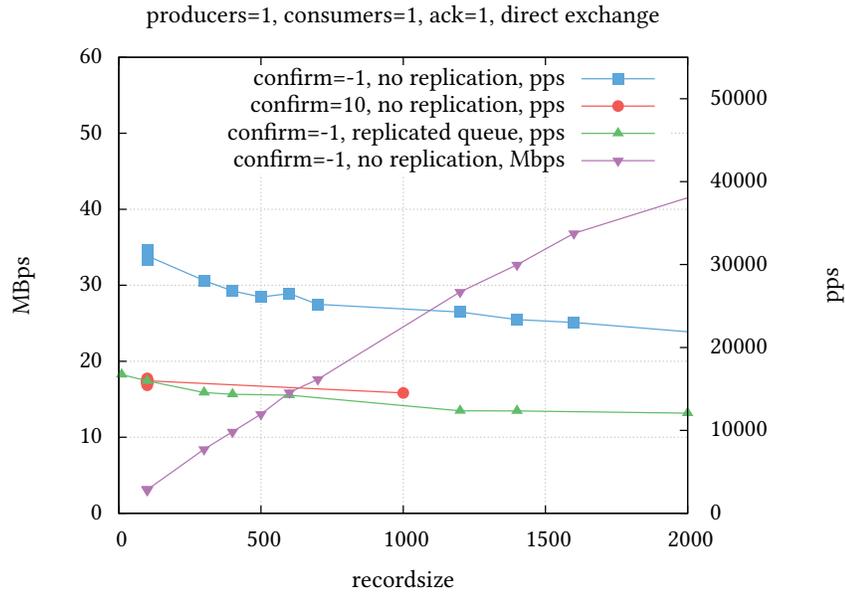

(a) RabbitMQ

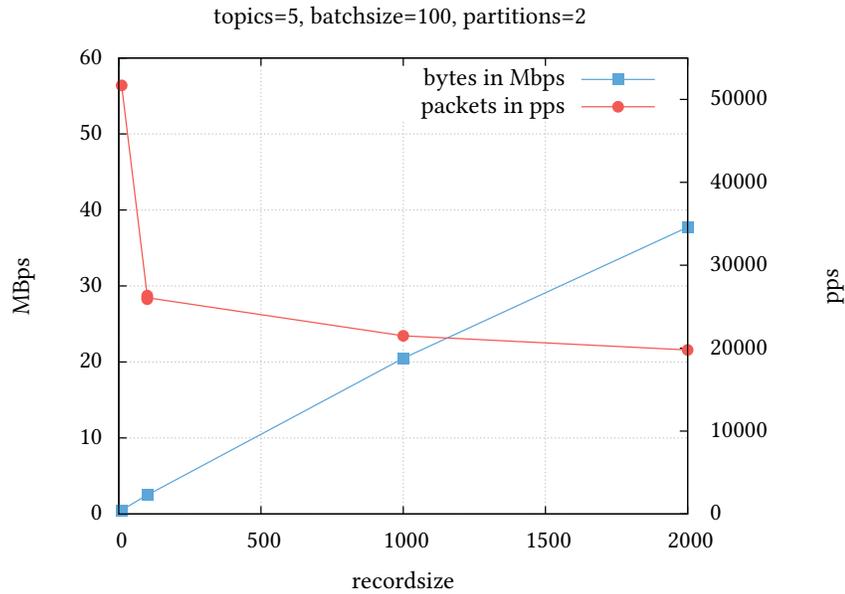

(b) Kafka

Fig. 4. Throughput as function of record size

*5.1.2 Latency in At Least Once Mode.* RabbitMQ latency is not really impacted by switching to a higher level of reliability: the packet will be written out to disk but since it is also available in memory this does not impact how fast it can be consumed.

For Kafka, the latency increases in case of replication since Kafka only delivers messages to consumers when they are acknowledged by a quorum of the active replicas (this is needed since Kafka does not enforce an fsync per packet on any of the replicas, so a Kafka packet is only protected by the fact it is kept by multiple machines).

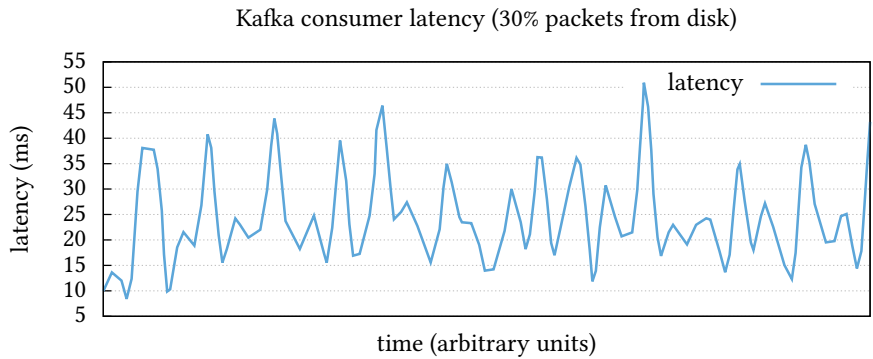

Fig. 5. Kafka cache miss latency [8]

**Summary.** In case of RabbitMQ, up to medium level of load, the latency for both at most once and at least once modes is below 10 ms.

In case of Kafka, on the other hand, if it can read from OS cache, its latency for at most once mode is below 10 ms, and about twice as large for the at least once mode. However, when it needs to read from disk, its latency can grow by up to an order of magnitude to around 100 ms.

### 5.2 Throughput Results

RabbitMQ parallelization inside the node boils down to multithreading with Erlang actors, and parallelization across nodes can be either traffic partitioning across standalone nodes or traffic distribution inside clusters. Kafka parallelization inside the node is also due to multithreading (influenced by producer and consumer count). Kafka parallelization across nodes is due to partitions, see Section 3.1. The performance tests that we have run only consider a single node, so they need to be adjusted with a suitable factor expressing the external parallelization.

*5.2.1 Throughput in At Most Once Mode.* Throughput for RabbitMQ is optimal if the broker is configured to allow an unlimited number of unconfirmed publishes ($confirm == -1$).

Figure 4a shows the impact of record size (in bytes) on throughput for a single RabbitMQ node. In these figures, pps stands for packets per second. As is to be expected, throughput decreases for larger packets (in addition to the packet switching effort which does not depend on the size, the byte copying operation scales linearly with the record size). Performance is optimal in the case of unlimited outstanding confirmed publishes. Replication lowers the throughput.

As reported in [24], using a clustering setup on Google Compute Engine consisting of 32 node, using 186 queues, 13000 consumers and producers and a load balancer in front, RabbitMQ was able to sustainably handle over 1.3M pps.

While measuring throughput of Kafka, the three important factors are: record size, partition count, and topic count. We have conducted different experiments to investigate the impact of each of these factors. Our findings are described below.

Figure 4b shows how the record size influences the throughput (in MBps or pps) of Kafka. The throughput in packets curve has a similar shape as what we found for RabbitMQ. When we plot the throughput in bytes per unit of time, we observe an almost linear relation to the record size: copying packets in Kafka is the dominant operation.

Figure 6 shows how the throughput of Kafka is impacted by the number of topics. It is important to point out that all these topics are active topics, each served by an individual producer. Based on this graph, it is a linear relation, however, the linearity in topics/producers has a hard limit of about 8 producers in our experimental setup. Machine utilization at that peak was Kafka 3.25 cores, all producers 10 cores, all consumers 4 cores and about 6 cores idle. With a higher topic count, performance diminishes.

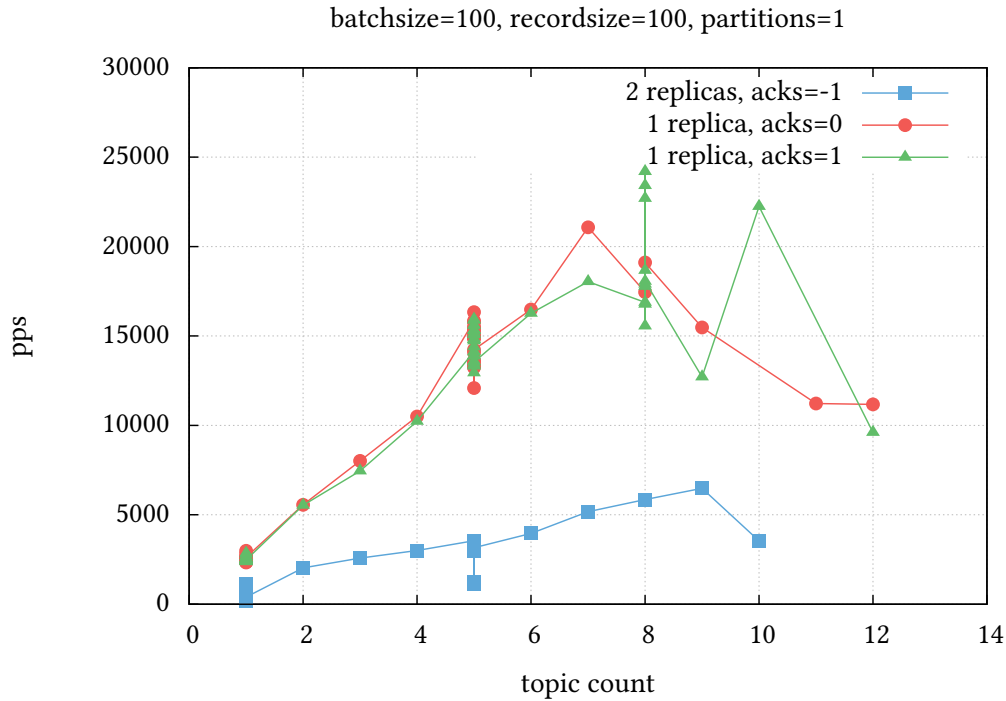

Fig. 6. Kafka throughput as function of topic count

Kafka up to about v0.8.2 was not designed to handle a large number of topics (with a hard limit around 10000 due to Zookeeper limits) as is evident from test results in [14]. These tests differ from ours in 2 ways: they use a partition count that maximizes throughput and not every topic was loaded by test generators. The results show a rapid deterioration in throughput for topic counts between 100 and 1000. Note that both in our experiments and the experiments from [14], setting up a lot of topics or a lot of partitions (a few hundred) led to frequent crashes of the control plane logic.

Figure 7 shows the throughput of Kafka as a function of partition counts. Its slope tapers off at about 10 (not due to core utilization, presumably by disk cache / driver logic resource contention) and the curve peaks at 200 in our experimental setup. This peak will occur elsewhere on systems with different core / DRAM / performance specs, so evidently, determining the partition count will be one of the most important configuration jobs of a Kafka installation.

In [34], Wyngaard reports on an experiment at NASA JPL for very high record sizes (10MB). A maximum of 6.49Gbps for throughput measured in bytes was found on a configuration with 3 producers, 3 consumers, single Kafka instance. More producers or more Kafka instances reduced this performance.

The observation that increasing the number of partitions beyond a certain point does not help to increase total throughput anymore could be due to the fact that the batchsize defined at the producer is split over the number of partitions.

*5.2.2 Throughput in At Least Once Mode.* For RabbitMQ, at least once mode implies writing packets to disk. The read will still happen from memory, as long as the packet in memory has not been trashed.

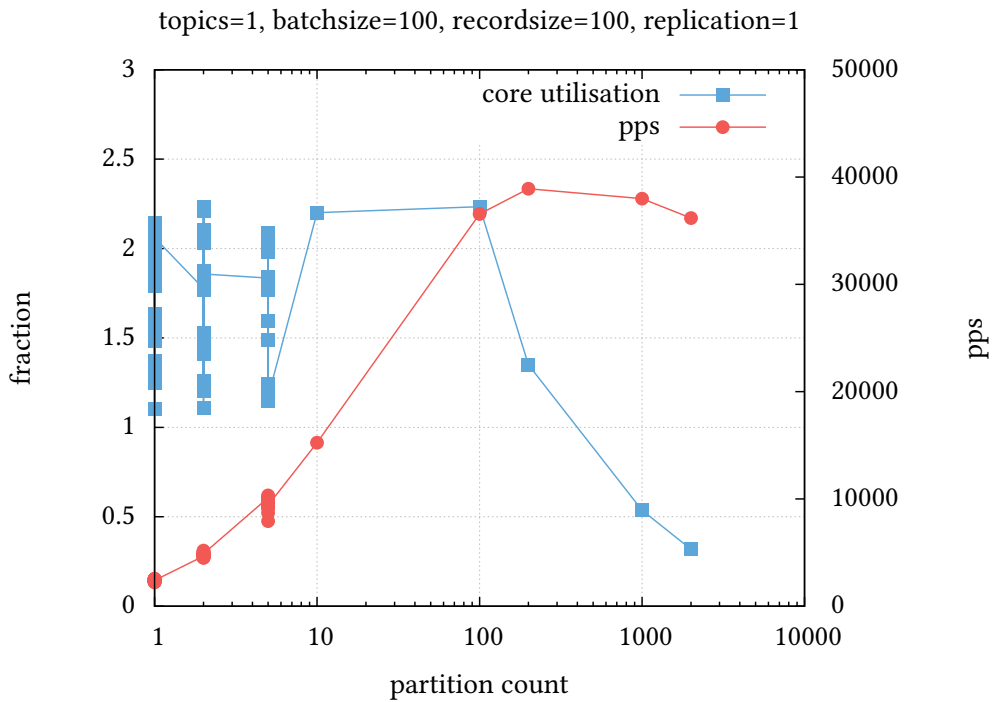

Fig. 7. Kafka throughput as function of partition count

Using producer batches is supported by AMQP, and RabbitMQ allows individual ACK/NACK responses per message in the batch. Together with the insertion sort on the queues, this ensures a producer can pipeline several batches without the need to wait for ACKS and still be sure about message order in the queues.

Switching on mirrored queues will have a negative impact on this throughput since the ACK now needs to be guaranteed from all the replicated queues.

Referring to Figure 4a above, for at least once delivery scenarios, RabbitMQ's throughput drops by 50% compared to the best effort scenario.

The case of Kafka is more complicated to analyze. Kafka assumes packets are written in batches, and will always write them to disk. In its at most once mode, Kafka can define a window (either in terms of number of packets or in terms of time, "log.flush.interval.messages" and "log.flush.interval.ms") of outstanding messages that have not yet been fsynced to disk media. This implies that on a system crash these messages will be lost.

The only way to get truly reliable message delivery with Kafka is running it in a mode where acknowledges are sent only if a batch has been written to the disk medium, or has been received by a quorum of replicas. This effectively slows down the producer to wait for a confirmation of a batch before the next batch can be written, introducing a round trip delay which is not required in RabbitMQ due to its selective ACK/NACK mechanism and reordering logic. If Kafka is running with redundant nodes, the client has to wait for a quorum ACK which will even take longer (2 round trips instead of 1). Our experiments show a decrease in performance due to replication of about 3 to 4 for $topics = 5, batchsize = 100, partitions = 5, replication = 2$. A similar experiment, reported in [23], generated slightly better results. Lastly, results from [22] indicate that the impact of replication factor is visible only when acknowledgments from all replicas are taken into account. For a 2 replica system, performance drops to half for larger batches, to a third for single messages.

$$\frac{|producers|}{U_{routing}+size*U_{byte}}$$

(a)

|  | $U_{routing}$ | $U_{byte}$ | Mean Error |
|---|---|---|---|
| no replication | $3.24e-5$ | $7.64e-9$ | 3% |
| replicated queue | $6.52e-5$ | $8.13e-9$ | 4.5% |

(b)

Table 3. Modeling the throughput of RabbitMQ: (a) suggested function (b) fitted values

$$\frac{|producers|*|partitions|}{U_{routing}+|topics|*U_{topics}+effective\_size^{0.5}*U_{byte}}$$

(a)

|  | $U_{routing}$ | $U_{topics}$ | $U_{byte}$ | Mean Error |
|---|---|---|---|---|
| acks = 0, rep. = 0 | $3.8e-4$ | $2.1e-7$ | $4.9e-6$ | 30% |
| acks = 1, rep. = 0 | $3.9e-4$ | $9.1e-8$ | $1.1e-6$ | 30% |
| acks = -1, rep. = 2 | $9.4e-4$ | $7.3e-5$ | $2.9e-5$ | 45% |

(b)

Table 4. Modeling the throughput of Kafka: (a) suggested function (b) fitted values

We conclude that Kafka's throughput in at least once mode decreases by 50% to 75% compared to the best effort scenario.

**Analysis and Summary.** There are various factors contributing to the overall throughput of RabbitMQ and Kafka. In order to simplify the analysis and summarize the results reported in this section, we use a curve-fitting approach: for each system we propose a simple first order function to model its throughput characteristics based on its inputs as well as important architecture/implementation details. Furthermore, in each case we apply a minimum error procedure between the proposed function and our empirical data.

The resulting functions and fitted values for RabbitMQ and Kafka are depicted in Table 3 and Table 4, respectively. Note that in this tables, U stands for "(processor cycle) utilization" and effective_size (in case of Kafka) is the maximum of the batch size and the record size used by the producers.

The proposed function to model the throughput behavior of RabbitMQ is shown in Table 3. Due to the Actor Model of Erlang, the total throughput per RabbitMQ node scales linearly with the number of producers, hence the producer factor in the formula. The producers factor, however, is only linear up to a value governed by how Erlang distributes its lightweight processes over cores, in our measurements the pps already saturated at the value corresponding to *producers == 2* for a single node and single queue.

Part (b) of Table 3 shows the fitted values of the proposed function for the graph in Figure 4a. The mean error is significantly low. Another fit on the measurements published in [25] gives very similar results.

There are a few important remarks here: (i) there is a small influence on the outstanding published packets parameter. If we change this from 10 to 100, for 100 bytes packets the throughput increases to 20Kpps, (ii) all these results are for a direct exchange. A topic exchange has more complex routing logic (for 100 byte packets, the throughput lowers to 19Kpps).

In summary, we can conclude that RabbitMQ is mainly constrained by routing complexity (up till frame sizes of a few 1000 bytes, at which time packet copying becomes non-negligible), which is the reason why we prefer to express RabbitMQ performance in packets per unit of time.

The proposed function to model the throughput behavior of Kafka is shown in Table 4. The "topics" parameter counts the number of configured topics on the Kafka broker. It is worth noting that for Kafka, we get the best fit if we put a 0.5 exponent, which might be related to the power law of cache misses.

For $producers = 5, size = 4000, partitions = 10$ our estimation predicts 85 Kpps. On a slightly more powerful processor architecture (faster memory, twice the cache size), [16] reports 140 Kpps for a similar test configuration.

From these parameters, it becomes evident that it is more appropriate to express Kafka throughput in bytes, since $U_{byte}$ is dominant even for small frames.

Finally, the error rate level in case of Kafka is not as low as that of RabbitMQ. Two potential causes for these variations are: (i) Kafka relies on OS level caching of disk access, which is a complex hidden subsystem that cannot be accurately modeled or even controlled and is shared across everything that runs on the machine (ii) Kafka runs on the JVM, which has much more variability [26] than an Erlang VM due to unsophisticated locking mechanisms and the garbage collection process.

## 6 DISTINCT FEATURES

In the previous sections, we looked at the common features that Kafka and RabbitMQ share. However, these two systems also come with their own distinct features. Knowledge of such features might be an important factor while making the decision to choose one of the two. Hence, below, we give a short summary of such features.

### 6.1 Features Unique to Kafka

*6.1.1 Long Term Message Storage.* Kafka stores its messages on disk. Purging of messages is done automatically and configured per topic. Messages are purged either after a retention time or when the topic's disk quota has been exceeded.

*6.1.2 Message Replay.* Since Kafka keeps no state about consumers and messages can be stored long term, consumers can easily replay messages when needed. This can be a very useful feature for the fault tolerance of the downstream systems.

*6.1.3 Kafka Connect.* Kafka Connect is a framework for scalable and reliable streaming of data between Apache Kafka and other systems. It makes it simple to quickly define connectors that move large collections of data into and out of Kafka.

*6.1.4 Log Compaction.* Kafka's log compaction feature ensures that it will always retain at least the last known value for each message key within the log of data for a single topic partition. This can be particularly useful in the use cases that are based on change feeds (defined in Section 7).

The Kafka ecosystem offers libraries and tools that provide additional functionality on top of Kafka as pub/sub system. A notable example is **Kafka Streams** which is briefly explained in Section 7.1.5. A detailed description of these capabilities is beyond the scope of this paper.

### 6.2 Features Unique to RabbitMQ

*6.2.1 Standardized Protocol.* RabbitMQ is, in essence, an open-source implementation of AMQP, a standard protocol with a highly-scrutinized design. As such, it enjoys a higher level of interoperability and can easily work with (and even be replaced by) other AMQP-compliant implementations.

*6.2.2 Multi-protocol.* In addition to AMQP, RabbitMQ supports a few other industry standard protocols for publishing and consuming messages, most notably MQTT (a very popular choice in the IoT community) and STOMP. Hence, in settings with mixed use of protocols, RabbitMQ can be a valuable asset.

*6.2.3 Distributed Topology Modes.* RabbitMQ, in addition to clustering, also supports federated exchanges which is a good match for Wide-area deployment with less-reliable network connections[6]. Compared to Clustering, it has a lower degree of coupling. A very useful feature of the federated exchanges is their on-demand forwarding. Furthermore, through its Shovel mechanism, RabbitMQ provides another convenient and easy way to chain brokers/clusters together.

*6.2.4 Comprehensive Management and Monitoring Tools.* RabbitMQ ships with an easy-to-use management UI that allows user to monitor and control every aspect of the message broker, including: (i) connections, (ii) queues, (iii) exchanges, (iv) clustering, federation and shoveling, (v) packet tracing, (vi) resource consumption. Together, these offer excellent visibility on internal metrics and allow for easy test and debug cycles.

*6.2.5 Multi-tenancy and Isolation.* RabbitMQ implements the notation of Virtual Hosts which is defined by AMQP to make it possible for a single broker to host multiple isolated environments (i.e. logical groups of entities such as connections, exchanges, queues, bindings, user permissions, policies, etc).

*6.2.6 Consumer Tracking.* At queue level, it keeps state, and knows exactly what consumers have consumed what messages at any time.

*6.2.7 Disk-less Use.* RabbitMQ does not require disk space to route packets, if persistence is not a requirement. This makes it a good choice for embedded applications and restricted environments. In fact, RabbitMQ has been successfully deployed on Raspberry Pi [6].

*6.2.8 Publisher Flow Control.* RabbitMQ can stop publishers from overwhelming the broker in extreme situations. This can be used in a flow control scenario when deletion of messages is not acceptable.

*6.2.9 Queue Size Limits.* A queue can be limited in size. This mechanism can help in a flow control scenario when deletion of messages is acceptable.

*6.2.10 Message TTL.* A message can be given a "Time To Live". If it stays beyond that time in any queue, it will not be delivered to the consumer. This makes a lot of sense for realtime data that becomes irrelevant after a specific time. The TTL can be attached to a queue at creation time, or to individual messages at the time of publishing.

## 7 PREFERRED USE CASES
### 7.1 Best Suited for Kafka

*7.1.1 Pub/Sub Messaging.* Kafka can be a good match for the pub/sub use cases that exhibit the following properties: (i) if the routing logic is simple, so that a Kafka "topic" concept can handle the requirements, (ii) if throughput per topic is beyond what RabbitMQ can handle (e.g. event firehose).

*7.1.2 Scalable Ingestion System.* Many of the leading Big Data processing platforms enable high throughput processing of data once it has been loaded into the system. However, in many cases, loading of the data into such platforms is the main bottleneck. Kafka offers a scalable solution for such scenarios and it has already been integrated into many of such platforms including Apache Spark and Apache Flink, to name a few.

---

[6]Recent versions of Kafka have a notion of federation, but more in the sense of cross-datacenter replication.

*7.1.3 Data-Layer Infrastructure.* Due to its durability and efficient multicast, Kafka can serve as an underlying data infrastructure that connects various batch and streaming services and applications within an enterprise.

*7.1.4 Capturing Change Feeds.* Change feeds are sequences of update events that capture all the changes applied on an initial state (e.g. a table in database, or a particular row within that table). Traditionally, change feeds have been used internally by DBMSs to synchronize replicas. More recently, however, some of the modern data stores have exposed their change feeds externally, so they can be used to synchronize state in distributed environments. Kafka's log-centric design, makes it an excellent backend for an application built in this style.

*7.1.5 Stream Processing.* Starting in Kafka version 0.10.0.0, a light-weight stream processing library called Kafka Streams is available in Apache Kafka to perform stateful and fault-tolerant data processing. Furthermore, Apache Samza, an open-source stream processing platform is based on Kafka.

**7.2 Best Suited for RabbitMQ**

*7.2.1 Pub/Sub Messaging.* Since this is exactly why RabbitMQ was created, it will satisfy most of the requirements. This is even more so in an edge/core message routing scenario where brokers are running in a particular interconnect topology.

*7.2.2 Request-Response Messaging.* RabbitMQ offers a lot of support for RPC style communication by means of the correlation ID and direct reply-to feature, which allows RPC clients to receive replies directly from their RPC server, without going through a dedicated reply queue that needs to be set up.

Hence, RabbitMQ, having specific support to facilitate this usecase and stronger ordering guarantees, would be the preferred choice.

*7.2.3 Operational Metrics Tracking.* RabbitMQ would be a good choice for realtime processing, based on the complex filtering the broker could provide.

Although Kafka would be a good choice as an interface to apply offline analytics, given that it can store messages for a long time and allows replay of messages. Throughput per topic could be another criterion to decide.

*7.2.4 Underlying Layer for IoT Applications Platform.* RabbitMQ can be used to connect individual operator nodes in a dataflow graph, regardless of where the operators are instantiated. A lot of the features of RabbitMQ directly cover platform requirements: (i) sub 5ms latency for the majority of the packets, throughput up to 40Kpps for single nodes, (ii) excellent visibility on internal metrics and easy test and debug cycles for dataflow setup through the web management interface, (iii) support for the MQTT protocol, (iv) sophisticated routing capability allows to expose packet filters as part of an associated data processing language, and (v) the possibility to handle a very large number of streams that all have rather small throughput requirements, with a large number of applications all interested in different small subsets of these streams.

*7.2.5 Information-centric Networking.* This is essentially a game on the capabilities of the architecture to intelligently route packets. Therefore, RabbitMQ would be the preferred choice, maybe even with a specific exchange that understands the link between routing key and destination. The geographic routing described in [12] is an example.

## 7.3 Combined Use

There are a number of requirements that cannot be covered solely by either RabbitMQ or Kafka, and where a combination of both is the best option.

Two common options for chaining these two systems are the following:

- Option 1: RabbitMQ, followed by Kafka. This is a good choice if RabbitMQ would be the best architectural choice, but some streams need long term storage. By putting RabbitMQ first, stronger latency guarantees can be offered. It also allows fine-grained selection of what streams need to go to long term storage, preserving disk resources.
- Option 2: Kafka, followed by RabbitMQ. This is a good choice if the throughput for the whole system is very high, but the throughput per topic is within the bounds of what a single node RabbitMQ broker can handle. By putting a RabbitMQ node behind a Kafka topic stream, all the complex routing capabilities of RabbitMQ can be combined with the complementary features of Kafka.

The AMQP-Kafka Bridge [27] can facilitate the interactions between RabbitMQ and Kafka.

Alternatively, RabbitMQ and Kafka can just be put in parallel, both processing the same input streams. This is more likely to happen in a scenario where two existing architectures are merged, and one was using Kafka while the other was using RabbitMQ.

**Determination Table.** So far we have considered specific use cases whose requirements are best satisfied by Kafka, RabbitMQ or a combination of both. In order to make these recommendations applicable to other use cases, we propose a determination table (depicted in Table 5). Each row in the table shows a set of features, and the architectural choice that corresponds to this set. This table obviously oversimplifies the decision to take - architects are advised to consider all dimensions of the problem as discussed in Sections 4, 5 and 6 before coming to a conclusion.

## 8 CONCLUSION

Kafka and RabbitMQ are two popular implementations of the pub/sub interaction paradigm. Despite commonalities, however, these two systems have different histories and design goals, and distinct features. RabbitMQ is an efficient implementation of the AMQP protocol, that offer flexible routing mechanism, using the exchanges/binding notions. It is much closer to the classic messaging systems. For example, it takes care of most of the consumption bookkeeping, its main design goal is to handle messages in memory, and its queue logic is optimized for empty-or-nearly-empty queues. Kafka, on the other hand, is designed around a distributed commit log, aiming at high-throughput and consumers of varying speeds. To that end, it has departed from the classic principles of messaging systems in a few ways: extensive use of partitioning at the expense of data order, its queues are logical views on persisted logs, allowing replayability, but manual retention policies. Furthermore, it also applies a number of very effective optimization techniques, most notably, aggressive batching and reliance on persistent data structures and OS page cache.

In this paper, we established a comparison framework to help position Apache Kafka and RabbitMQ w.r.t. each other, both quantitatively and qualitatively.

In terms of latency, both systems are capable of delivering low-latency results (i.e., mean/median of around 10 ms). In case of RabbitMQ, the difference between at most once and at least once delivery modes is not significant. For Kafka, on the other hand, latency increases about twice as large for the at least once mode. Additionally, if it needs to read from disk, its latency can grow by up to an order of magnitude.

| predictable latency? | complex routing? | long term storage? | very large throughput per topic? | packet order important? | dynamic elasticity behavior? | system throughput? | at least once? | high availability? | |
|---|---|---|---|---|---|---|---|---|---|
| N | N | * | * | N | N | XL | N | N | Kafka with multiple partitions |
| N | N | * | * | N | N | XL | Y | Y | Kafka with replication and multiple partitions |
| N | N | * | * | Y | N | L | N | N | single partition Kafka |
| N | N | * | * | Y | N | L | Y | Y | single partition Kafka with replication |
| * | * | N | N | * | * | L | * | N | RabbitMQ |
| * | * | N | N | * | * | L | * | Y | RabbitMQ with queue replication |
| * | * | Y | N | * | * | L | * | * | RabbitMQ with Kafka long term storage |
| N | Y | * | * | N | N | XL | N | * | Kafka with selected RabbitMQ routing |

[1] Y - feature required, N - feature not required, * - wildcard, replaces two rows that are identical but in this feature, one with Y and one with N

[2] L(arge), (e)X(tra)L(arge), see 5.2 for some more quantitative throughput figures

Table 5. RabbitMQ and/or Kafka?

In terms of throughput, in the most basic set up (i.e. on a single node, single producer/channel, single partition, no replication) RabbitMQ's throughput outperforms Kafka's. Increasing the Kafka partition count on the same node, however, can significantly improve its performance, demonstrating its superb scalability. Increasing the producer/channel count in RabbitMQ, on the other hand, could only improve its performance moderately.

Both Kafka and RabbitMQ can scale further by partitioning flows over multiple nodes. In RabbitMQ, this requires additional special logic, such as Consistent Hash Exchange [3] and Sharding Exchange [28]. In Kafka this comes for free. Finally, replication has a drastic impact on the performance of both RabbitMQ and Kafka and reduces their performance by 50% and 75%, respectively.

While efficiency aspects are very important, architects are strongly advised to consider all other dimensions of the problem as discussed in Sections 4 (qualitative comparison of common features beyond performance) and 6 (distinct features) before coming to a conclusion. The study reported in [26] which was conducted in the context of a real-world application can serve as a good example.

Further, as described in section 7, such choice does not have to be an exclusive one and a combination of both systems might be the best option.